\documentclass[aip,jcp,reprint]{revtex4-1}
\usepackage[T1]{fontenc}
\usepackage{amsmath}
\usepackage[latin1]{inputenc}
\usepackage{graphicx}

\usepackage{epsfig}

\usepackage{color}

\begin{document}
\title{Phase diagrams of binary mixtures of patchy colloids with distinct numbers and types of patches: The empty fluid regime}

\author{Daniel de las Heras}
\email{delasheras.daniel@gmail.com}
\affiliation{Centro de F\'{i}sica Te\'orica e Computacional da Universidade de Lisboa, Avenida Professor Gama Pinto 2, P-1749-016, Lisbon, Portugal}

\author{Jos\'e Maria Tavares}
\email{josemaria.castrotavares@gmail.com}
\affiliation{Instituto Superior de Engenharia de Lisboa, Rua Conselheiro Em\'{\i}idio Navarro, P-1590-062 Lisbon, Portugal, and Centro de F\'{i}sica Te\'orica e Computacional da Universidade de Lisboa, Avenida Professor Gama Pinto 2, P-1749-016, Lisbon, Portugal}

\author{Margarida M. Telo da Gama}
\email{margarid@cii.fc.ul.pt}
\affiliation{Departamento de F\'{\i}sica, Facultade de Ci\^encias da Universidade de Lisboa, Campo Grande, P-1749-016, Lisbon, Portugal, and Centro de F\'{i}sica Te\'orica e Computacional da Universidade de Lisboa, Avenida Professor Gama Pinto 2, P-1749-016, Lisbon, Portugal}

\date{\today}
\begin{abstract}

We investigate the effect of distinct bonding energies on the onset of criticality of low functionality fluid mixtures. We focus on mixtures of particles with two and three patches as this includes the mixture where 'empty' fluids were originally reported. In addition to the number of patches, the species differ in the type of patches or bonding sites. For simplicity, we consider that the patches on each species are identical: one species has $3$ patches of type $A$ and the other $2$ patches of type $B$. We have found a rich phase behaviour with closed miscibility gaps, liquid-liquid demixing and negative azeotropes. Liquid-liquid demixing was found to preempt the 'empty' fluid regime, of these mixtures, when the $AB$ bonds are weaker than the $AA$ or $BB$ bonds. By contrast, mixtures in this class exhibit 'empty' fluid behaviour when the $AB$ bonds are stronger than at least one of the other two. Mixtures with bonding energies $\epsilon_{BB}=\epsilon_{AB}$ and $\epsilon_{AA}<\epsilon_{BB}$, were found to exhibit an unusual negative azeotrope. 

\end{abstract}

\maketitle

\section{Introduction \label{introduction}} 

In the last twenty years the study of colloidal phase diagrams, based on spherically symmetric particle interactions, revealed a rich phenomenology including new crystal phases, gelation and glass transitions \cite{pusey1,zaccarelli1}. The phase behaviour of binary mixtures is richer and colloidal binary mixtures are expected to have a wider range of applications in technology as well as in biology. Examples include novel candidate photonic crystals, synthesized by van Blaaderen and co-workers, through fine control of the colloidal charges \cite{Leunissen1,Hynninen1}, and binary mixtures of eye-lens proteins, studied by Schurtenberger et al, where attractive unlike (spherical) interactions were found to be crucial to stabilize the mixture, a mechanism relevant in the prevention of cataract formation \cite{PhysRevLett.99.198103}. 

Nowadays colloidal particles can be synthesised in a range of shapes and their surfaces may be functionalized in a variety of ways \cite{intro1}, with the result that the particle interactions become directional or 'patchy'. The primitive model of patchy colloids consists of hard-spheres with $f$ patches on their surfaces. Patchy particles attract each other if and only if two of their patches overlap. The attraction between particles is short ranged and anisotropic: The patches act as bonding sites and promote the appearance of well defined clusters, whose structure and size distribution depend on the properties of the patches ($f$ and the bonding energy) and on the thermodynamic conditions (density and temperature). 

Sciortino and co-workers established that $f$, the number of patches or bonding sites per particle, is the key parameter controlling the location of the liquid-vapour critical point \cite{PhysRevLett.97.168301,bianchi2008}. They showed that, for low values of $f$ (approaching 2), the phase separation region is drastically reduced, and low densities and temperatures can be reached without encountering the phase boundary. These low density ('empty') phases were shown to be network liquids, suggesting that, on cooling, patchy particles could assemble into a glassy state of arbitrary low density (a gel). 

Remarkably, the results of the simulations of patchy colloidal particles are well described by classical liquid state theories: Wertheim's first order perturbation theory \cite{wertheim1,*wertheim2,*wertheim3,*wertheim4} predicts correctly the equilibrium thermodynamic properties; Flory-Stockmayer \cite{flory1,*stock1,*flory2} theories of polymerization describe quantitatively the size distributions of the clusters of patchy particles, including the appearance of network (percolated) fluids. 

In subsequent work we addressed, explicitly, the interplay of the entropy of mixing and the entropy of bonding in the phase behaviour of models of binary mixtures of patchy particles. We focused on mixtures of particles which differ in the number of patches or functionality only. We found that, within this class of mixtures, the difference between the functionality of the particles is the key parameter controlling the phase equilibria of the system. In particular, if one species has more than twice the number of bonding sites of the other, a phase transition between two network fluids appears and the topology of the phase diagram may change. The miscibility at high pressures is also controlled by this difference and closed miscibility gaps are always present when the difference, between the functionality of the particles, is greater than one \cite{mixtures1}. The conditions required to realize 'empty' phases were shown to differ from those proposed by Sciortino and co-workers, and no 'empty' liquids were found when the functionality of one of the species exceeds 4 \cite{mixtures1}.

The primitive model of patchy particles was generalized to include different types of patches ({\it i.e.} more than one bonding energy). This was achieved \cite{tavares1,*tavares4,tavares3} by considering a single component fluid where the particles have two A sites  (with bonding energy $\epsilon_{AA}$), and one B site (with bonding energy $\epsilon_{BB}$). Unlike sites also interact with bonding energy $\epsilon_{AB}$. This model allows a deeper understanding of the onset of criticality in low functionality systems: The detailed fashion in which the critical temperature vanishes, as the bonding energies decrease towards zero, depends on the order in which the limits are taken, which in turn determines the type of network that is formed. 'Empty' phases are low-density structured (percolated) fluids and the emergence of criticality is related to this structure.  Thus, the generalized model provides a reference system for the microscopic description of the competition between condensation and self-assembly of equilibrium structured fluids as the theoretical predictions based on Wertheim's theory have been confirmed, recently, by Monte Carlo simulations \cite{PhysRevE.81.010501,MMprl}.

In the present work, we proceed to investigate the effect of distinct bonding energies on the onset of criticality of low functionality fluid mixtures. We focus on mixtures of particles with two and three patches as this is the type of mixture where 'empty' fluids were originally reported. In addition to the number of patches the species may differ in the type of patches or bonding sites. For simplicity, we assume that the patches on each species are identical, {\it i.e.} we consider mixtures of particles with two $B$ patches and particles with three $A$ patches: $ 2_B-3_A$ mixtures.   

\begin{figure}
\epsfig{file=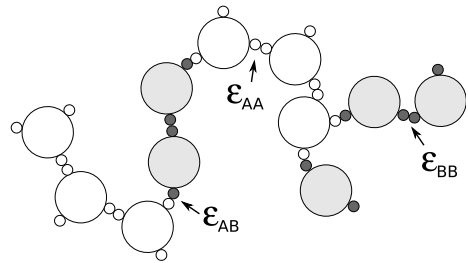,width=.7\linewidth,clip=}
\caption{Schematic representation of the mixture under consideration illustrating the different bonding energies.}
\label{fig1}
\end{figure}

As for pure fluids, we will use Wertheim's thermodynamic perturbation theory (in its extension to mixtures) \cite{Chapman:1057} and Flory-Stockmayer's theory of percolation (generalized to mixtures \cite{mixtures1}). The colloids are modelled by equisized hard spheres and we vary the strength of the bonding energies $\epsilon_{AA}$, $\epsilon_{BB}$ and $\epsilon_{AB}$ (Fig. \ref{fig1}). We recover the $2_A-3_A$ mixture investigated previously \cite{PhysRevLett.97.168301,mixtures1} when all bonding energies are equal. This is a type I mixture that is completely miscible above the critical pressure of the less volatile component. The 'empty' fluid regime is approached monotonically: The critical packing fraction and temperature decrease towards zero as the fraction of $2_A$ particles approaches one ({\it i.e.}, the average functionality decreases towards two). 

We found a rich phase behaviour with closed miscibility gaps, liquid-liquid demixing, negative azeotropes and, from a topological point of view, three different types of mixtures: type I, I-A and III \cite{Scott:49}. Liquid-liquid demixing is the mechanism which prevents this class of mixtures from exhibiting 'empty' fluid behaviour and this occurs when the $AB$ bonds are weaker than the $AA$ or the $BB$ bonds. By contrast, the mixtures exhibit 'empty' fluid behaviour when the $AB$ bonds are stronger than at least one of the other two. Furthermore, mixtures with bonding energies $\epsilon_{BB}=\epsilon_{AB}$, and $\epsilon_{AA}<\epsilon_{BB}$, exhibit an unusual (negative) azeotropic behaviour. 

We found an exception where the $AB$ bonds are stronger but the mixture fails to exhibit 'empty' fluid behaviour: the mixture with $\epsilon_{AB}=\epsilon_{AA}$ and $\epsilon_{BB}=0$. Near the critical point of this mixture there is demixing and two network fluids with finite packing fraction coexist, as the pressure vanishes. This behaviour resembles that found previously \cite{mixtures1} for $2_A-5_A$ mixtures where the entropy of bonding drives a demixing transition, which preempts the 'empty' fluid liquid-vapor phase transition.

The remainder of the paper is organised as follows. In section \ref{theory} we present the model, summarise Wertheim's theory for binary mixtures (Sec. \ref{wertheim}) and the Flory-Stockmayer theory of percolation for binary mixtures of patchy particles (Sec. \ref{percolation}). In section \ref{results} we present the results: phase diagrams (including percolation lines) and critical properties of several representative mixtures. Finally, in Sec. \ref{conclusions} we summarize our conclusions and suggest lines for future research.

\section{Model and theory \label{theory}}
We consider a binary mixture of $N_1$ and $N_2$ equisized hard spheres (HSs) with diameter $\sigma$. Particles of species $1$ are decorated with $2$ patches of type $B$, and particles of species $2$ with $3$ patches of type $A$: a $2_B-3_A$ binary mixture. The bonding sites are distributed on the particle surfaces in such a manner that two particles can form only one single bond, involving two bonding sites only, one in each particle. In addition, there is a minimum distance between the bonding sites to ensure that no sites are shaded by nearby bonds.
\subsection{Helmholtz free energy: Wertheim's thermodynamic perturbation theory \label{wertheim}}

A description of Wertheim's perturbation theory for pure fluids and mixtures can be found elsewhere \cite{wertheim1,*wertheim2,*wertheim3,*wertheim4,Chapman:1057}. Here we summarise the theory for the mixture under consideration (the theory for binary mixtures with an arbitrary number of patches of different types is discussed in \cite{mixtures1}).

The Helmholtz free energy per particle, $f_H$, is the sum of contributions from a reference system, a mixture of hard spheres ($f_{HS}$), and a perturbation due to the bonding interactions ($f_b$):
\begin{eqnarray}
f_{H}=F_H/N=f_{HS}+f_b,
\end{eqnarray}
where $N=N_1+N_2$ is the total number of particles. The HSs free energy may be written as the sum of ideal-gas and excess contributions: $f_{HS}=f_{id}+f_{ex}$. The ideal-gas free energy is given by
\begin{equation}
\beta f_{id}=\ln\eta-1+\sum_{i=1,2}x^{(i)}\ln(x^{(i)} {\cal V}_i),
\end{equation}
with $\beta=kT$ the inverse thermal energy, ${\cal V}_i$ the thermal volume and $x^{(i)}=N_i/N$ the number fraction of each species. $\eta=v_s\rho$ is the total packing fraction, $v_s=\pi/6\sigma^3$ the volume of a single particle and $\rho$ the total number density. The excess part, which includes the effect of the excluded volume, is approximated by the Carnahan-Starling equation of state \cite{carnahan:635} (note that both species have the same diameter):
\begin{equation}
\beta f_{ex}=\frac{4\eta-3\eta^2}{(1-\eta)^2}.
\end{equation}

The bonding free energy has two contributions: the bonding energy and an entropic term related to the number of ways of bonding two particles. In the framework of Wertheim's thermodynamic first-order perturbation theory it is given by \cite{Chapman:1057,mixtures1}
\begin{equation}
\beta f_b=x\left[2\ln X_B-X_B+1\right]+(1-x)\left[3\ln X_A-\frac32X_A+\frac32\right],\label{fb}
\end{equation}
where $x\equiv x^{(1)}$ is the composition of the mixture ($x^{(2)}=1-x$) and $X_\alpha$ is the probability that a site of type $\alpha=A,B$ is {\it not} bonded. The latter are related to the thermodynamic quantities through the law of mass action: 
\begin{eqnarray}
X_A=1-2\eta xX_BX_A\Delta_{AB}-3\eta(1-x)X_A^2\Delta_{AA}\nonumber\\
X_B=1-2\eta xX_B^2\Delta_{BB}-3\eta(1-x)X_AX_B\Delta_{AB} \label{mass}
\end{eqnarray}
The set of parameters $\Delta_{\alpha\gamma}$ characterise the bonds between sites $\alpha$ and $\gamma$. For simplicity, the interaction between bonding sites is modelled by square well potentials with depths $\epsilon_{\alpha\gamma}$. Assuming that all bonds have the same volume, $v_b$, and using the ideal-gas approximation for the pair correlation function of the reference HS fluid, we find \cite{mixtures1}
\begin{equation}
\Delta_{\alpha\gamma}=\frac{v_b}{v_s}\left[exp(\beta\epsilon_{\alpha\gamma})-1\right].
\end{equation}
As in previous works \cite{tavares1,*tavares4}, we set the bond volume to $v_b=0.000332285\sigma^3$. We consider mixtures where two of the three bonding energies, $\epsilon_{AA},\;\epsilon_{AB}$ and $\epsilon_{BB}$, are equal and vary the third. The two identical energies set the energy scale, $\epsilon$. In what follows we use scaled-bonding energies $\epsilon^*_{\alpha\gamma}=\epsilon_{\alpha\gamma}/\epsilon$.

We use the Gibbs free energy per particle ($g=p/\rho+f_H$, where $p$ is the pressure) to obtain the equilibrium properties of the mixture. At fixed values of the composition $x$, pressure $p$, and temperature $T$, $g$ is minimised with respect to the total density $\rho$, subject to the constraints imposed by the law mass action. We use a standard Newton-Raphson method to minimise $g$, and solve the law mass action simultaneously by a Powell hybrid method. 

A standard common tangent construction on $g(x)$ is used to determine the coexistence points (which is equivalent to imposing the equality of the chemical potentials in the coexisting phases) while mechanical and thermal equilibria are satisfied by fixing the pressure and the temperature, respectively.

Finally, the critical points are computed numerically by determining the states which satisfy the spinodal condition and the vanishing of the third-order derivative in the direction of largest growth \cite{lalmixtures}.

\subsection{Theory of percolation}\label{percolation}

\begin{figure}
\epsfig{file=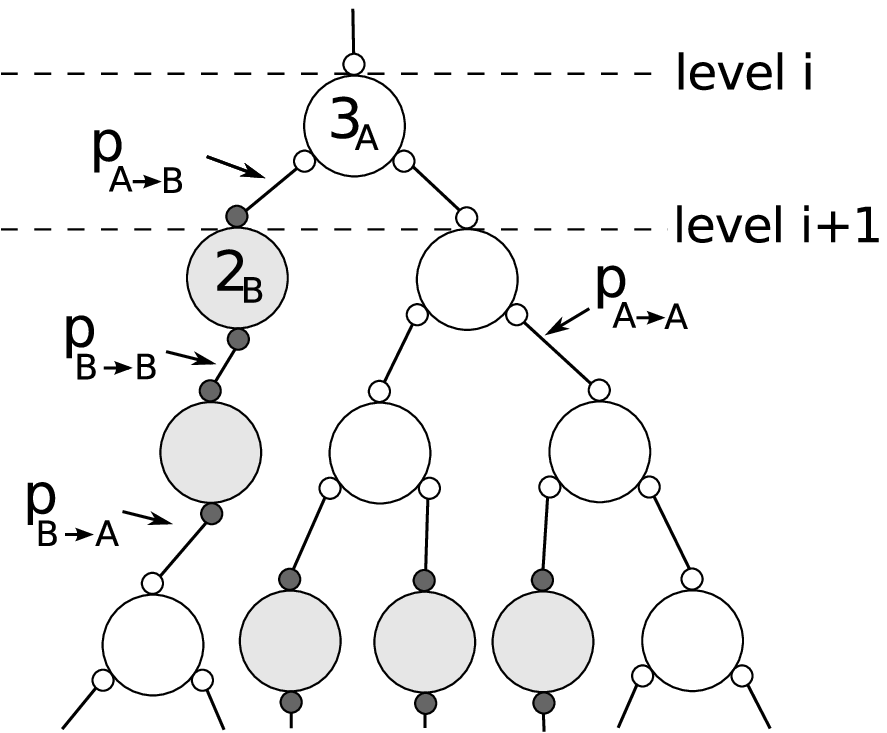,width=.7\linewidth,clip=}
\caption{Schematic representation of a tree-like cluster in a $2_B-3_A$ binary mixture. $p_{\alpha\rightarrow \gamma}$ is the probability of bonding a site $\alpha$ to a site $\gamma$.}
\label{fig2}
\end{figure}

Recently \cite{mixtures1} we generalised an extension of the Flory-Stockmayer random-bond percolation theory \cite{flory1,*stock1,*flory2} proposed by Tavares {\it et. al.} \cite{PhysRevE.81.010501} to binary mixtures with an arbitrary number of distinct bonding sites. Consider a tree-like cluster ({\it i.e.} with no loops) as schematically illustrated in Fig. \ref{fig2}. The number of bonded sites $\alpha$ at the level $i+1$ is related to the number of all types of bonded sites in the previous level through the recursion relations (see references \cite{mixtures1,PhysRevE.81.010501} for details): 
\begin{eqnarray}
n_{i+1,A}=2p_{A\rightarrow A}n_{i,A}+p_{B\rightarrow A}n_{i,B} \nonumber\\
n_{i+1,B}=2p_{A\rightarrow B}n_{i,A}+p_{B\rightarrow B}n_{i,B}\label{levels},
\end{eqnarray}
where $p_{\alpha\rightarrow \gamma}$ is the probability of bonding a site $\alpha$ to a site $\gamma$. Then, the probability of finding a bonded site $\alpha$ is
\begin{eqnarray}
P_A=p_{A\rightarrow A}+p_{A\rightarrow B}\nonumber\\
P_B=p_{B\rightarrow A}+p_{B\rightarrow B}\label{proba},
\end{eqnarray}
which may be related to the thermodynamic quantities through the law mass action, since
\begin{equation}
P_\alpha=1-X_\alpha,\quad\alpha=A,B.
\end{equation}
A term-by-term analysis of Eqs. (\ref{mass}) and (\ref{proba}) gives
\begin{eqnarray}
p_{A\rightarrow A}=(1-x)3\eta X_A^2\Delta_{AA}\nonumber\\
p_{A\rightarrow B}=x2\eta X_AX_B\Delta_{AB}\nonumber\\
p_{B\rightarrow B}=x2\eta X_B^2\Delta_{BB}\nonumber\\
p_{B\rightarrow A}=(1-x)3\eta X_AX_B\Delta_{AB}.
\end{eqnarray}

In order to find whether the system is percolated or not, we write the equations (\ref{levels}) in matrix form
\begin{equation}
\tilde n_i=\tilde T^i\tilde n_0,\label{progressions}
\end{equation}
where $\tilde n_i$ is a vector with components $n_{i,\alpha}$ and $\tilde T$ is a $2\times2$ square matrix with entries
\begin{equation}
\tilde T=
\begin{pmatrix}
2p_{A\rightarrow A} & p_{B\rightarrow A}  \\
2p_{A\rightarrow B} & p_{B\rightarrow B} 
\end{pmatrix},
\end{equation}
which may be diagonalized. The progressions defined by Eq. (\ref{progressions}) converge to $0$ if the largest (absolute value) of the eigenvalues of $\tilde T$ is less than unity. In other words, percolation occurs when $|\lambda_\pm|=1$ for any of the two eigenvalues $\lambda_\pm$ of $\tilde T$.

\section{Results \label{results}}

Before describing the results for $2_A-3_A$ mixtures we set the graphical code/s used in the figures. Phase diagrams are presented, in the temperature-composition plane at constant pressure, in Figs. \ref{fig3}, \ref{fig6}, and \ref{fig9}. In all the figures percolation lines are plotted as solid red lines: Below the percolation line the fluid is percolated in the sense that there is a non-zero probability of finding an infinite cluster. We call these networks fluids. The liquid side/s of the binodal lines is/are always percolated; binodal lines are depicted as black solid lines; shaded areas are two-phase regions; empty circles are critical points; black squares are azeotropic points.

The critical properties of the mixtures are analysed by means of pressure-temperature projections (Figs. \ref{fig4}, \ref{fig7} and \ref{fig10}) and critical temperature vs critical packing fraction representations (Figs. \ref{fig5}, \ref{fig8} and \ref{fig11}). In all cases: Solid lines depict the liquid-vapour  transition and empty circles the critical point of the pure fluid ($3_A$ fluid). Dashed and dotted lines are the critical points of the mixture: When the mixture exhibits 'empty' fluid behaviour, at low pressure, the critical line is dashed, otherwise it is shown dotted.

\subsection{$2_A-3_A$ mixture}

We start by reviewing the results for the mixture with identical bonding interactions, which was investigated by grand-canonical Monte Carlo simulations \cite{PhysRevLett.97.168301} and using the current theory \cite{mixtures1}. At low pressures, the pure $3_A$ fluid undergoes $LV$ liquid-vapour condensation, below the critical temperature. The transition involves two fluids with different densities and fraction of unbonded sites. By contrast, there is no $LV$ transition in the pure $2_A$ fluid, as the particles with two bonding sites can form linear chains only and the absence of branching prevents the fluids from condensing. For the mixture, the simulation and theoretical results indicate that the critical packing fraction decreases continuously towards zero as the pressure vanishes, exhibiting 'empty' liquid behaviour, that is, there is a fluid phase with arbitrary low packing fraction.

The temperature-composition phase diagrams are illustrated in Fig. \ref{fig3} panels ($b_1$) and ($b_2$) at two distinct pressures below $p_c^{(3)}$, the critical pressure of the pure $3_A$ fluid (the mixture is completely miscible above $p_c^{(3)}$). When $2_A$ particles are added the $LV$ phase transition shifts from $x=0$ ($3_A$ fluid) to finite values of the composition.  The two-phase region, which is always bounded by a lower critical point, increases as the pressure decreases. Near the $x=0$ axis the slope of the binodal is negative, indicating a decrease in the stability of the liquid phase as the fraction of $2_A$ particles increases. The addition of $2_A$ particles reduces the probability of branching, the mechanism which drives condensation, and consequently the region of stability of the liquid phase decreases.  

The $pT$ projection of the critical line (see Fig. \ref{fig4}) starts at the critical point of the less volatile fluid ($3_A$ fluid) and decreases monotonically to $p\rightarrow0$ and $T\rightarrow0$. Topologically this is a limiting case of type I mixtures, with no liquid-liquid demixing, according to the classification of van Konynenburg and Scott \cite{Scott:49}), where one of the fluids does not undergo $LV$ condensation ($2_A$ fluid).

In what follows, we consider generic $2_B-3_A$ mixtures ({\it i.e.} where species $1$ is decorated with $2$ patches of type $B$ and species $2$ with $3$ patches of type $A$) which are characterised by three bonding interactions: $\epsilon^*_{AA}$, $\epsilon^*_{BB}$ and $\epsilon^*_{AB}$. We proceed by setting, in turn, two of them equal and vary the third.

\subsection{$2_B-3_A$ mixtures: $\epsilon_{AA}=\epsilon_{AB}$}

\begin{figure*}
\epsfig{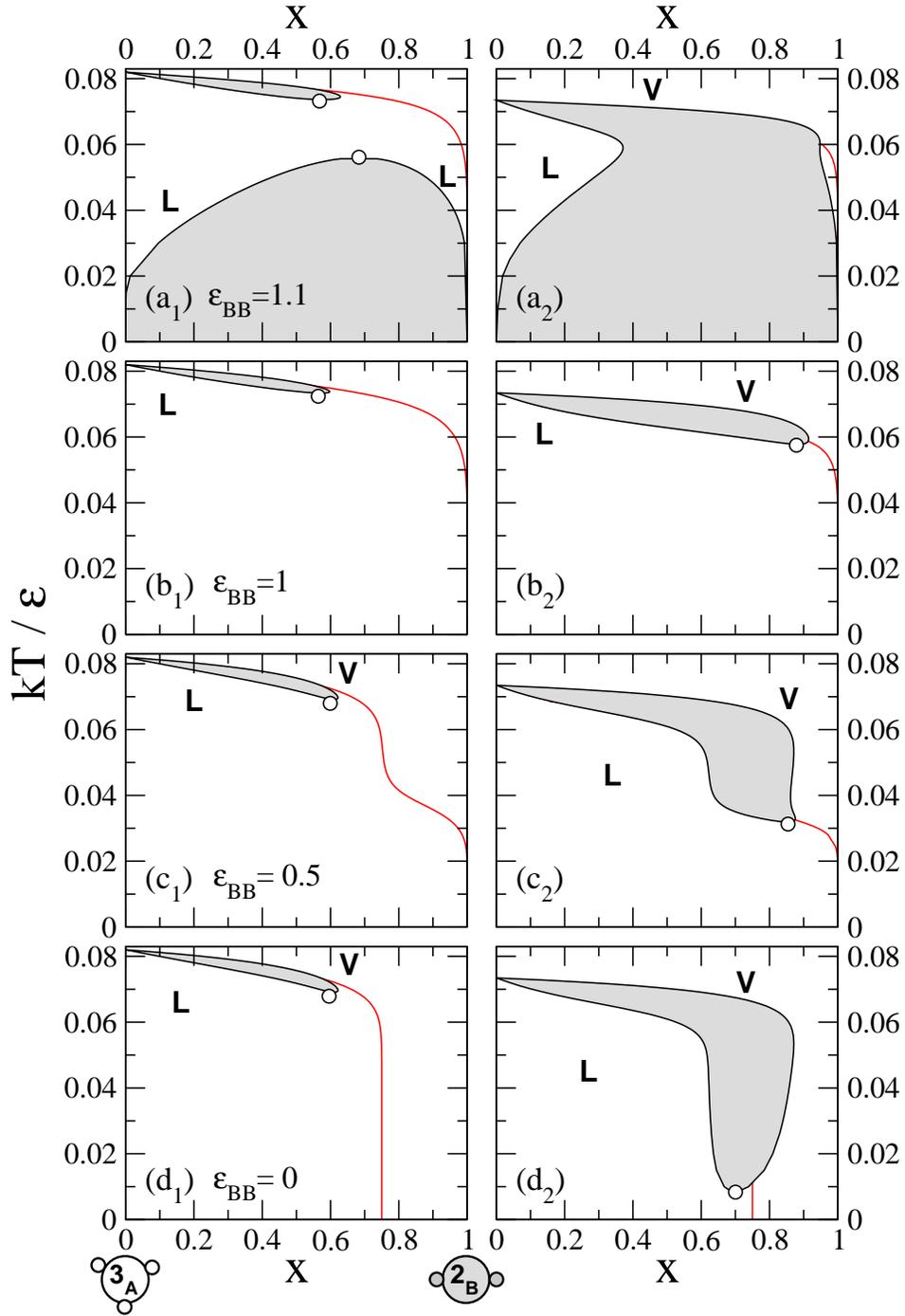}
\caption{Phase diagrams of mixtures of particles with $2$ patches of type $B$ (species $1$) and $3$ patches of type $A$ (species $2$) in the scaled-temperature vs composition ($x=x^{(1)}$) plane at constant pressure.  The bonding interactions are: $\epsilon^*_{AA}=\epsilon^*_{AB}=1$ and: row 1 ($a_1$ and $a_2$) $\epsilon^*_{BB}=1.1$, ($b_1$) and ($b_2$) $\epsilon_{BB}^*=1$, ($c_1$) and ($c_2$) $\epsilon^*_{BB}=0.5$, ($d_1$) and ($d_2$) $\epsilon^*_{BB}=0$. Left column: Pressure $p_1^*= p_1v_s/\epsilon=1.05\times10^{-4}$, right column: Pressure $p_2^*=p_2v_s/\epsilon=1.05\times10^{-5}$.}
\label{fig3}
\end{figure*}

We start by setting $\epsilon^*_{AA}=\epsilon^*_{AB}=1$. $Tx$ phase diagrams are depicted in Fig. \ref{fig3} at two different pressures: $p_1^*= p_1v_s/\epsilon=1.05\times10^{-4}$ (left column) and $p_2^*=1.05\times10^{-5}$ (right column), both below $p_c^{(3)}$, the critical pressure of the pure $3_A$ fluid. $pT$ projections are plotted in Fig \ref{fig4} and critical properties are illustrated in Fig. \ref{fig5}.

\subsubsection{$\epsilon^*_{BB}>1,\quad \epsilon^*_{AA}=\epsilon^*_{AB}=1$}\label{subsub1}

When $\epsilon^*_{BB}>1$ there is liquid-liquid demixing at low temperature for any value of the pressure. The demixing involves two structured (percolated) fluids: One rich in $3_A$ particles, and the other rich in $2_B$ particles. The phase separation is driven by the presence of $AB$ bonds as they reduce the number of $BB$ bonds which are energetically favourable. The demixing at low temperatures preempts the 'empty' fluid regime. 

We chose $\epsilon^*_{BB}=1.1$ as a representative example of this class of mixtures. At pressures higher than $p_c^{(3)}$ the $Tx$ phase diagram (not shown) consists in a demixing region bounded above by an upper critical point. Sightly below $p_c^{(3)}$ there are two regions of phase separation bounded by critical points (panel ($a_1$) of Fig. \ref{fig3}): $LL$ demixing at low temperatures and $LV$ at high temperatures. By reducing the pressure below a certain value ($p/p_c^{(3)}\approx0.27$) the two regions merge giving rise to a very large demixing region without critical points (panel ($a_2$) of Fig. \ref{fig3}). 

\begin{figure}
\epsfig{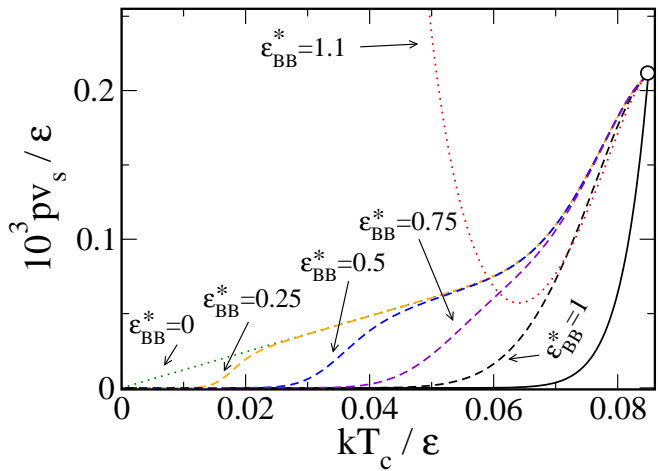}
\caption{Pressure-temperature projections of the phase diagrams of binary mixtures with $\epsilon^*_{AA}=\epsilon^*_{AB}=1$. The solid curve is the liquid-vapour transition of the pure $3_A$ fluid, which ends in a critical point (open circle). Dashed and dotted lines are the critical lines of the mixture.}
\label{fig4}
\end{figure}

The pure $2_B$ fluid does not undergo $LV$ condensation, and thus the phase diagram of these mixtures does not fit exactly the standard classification of van Konynenburg and Scott \cite{Scott:49}. Nevertheless, it may be understood as a limiting case of type III mixtures (see the $pT$ projection in Fig. \ref{fig4}). The critical line, which changes its character continuously from $LV$ to $LL$, starts at the critical point of the less volatile fluid, exhibits a minimum, and continues with negative slope to high pressures. This line is expected to intersect the solid phase boundaries that were not considered here. The topology of the phase diagram remains the same for other mixtures with $\epsilon^*_{BB}>1$, but changes in the critical line, such as the absence of the minimum, may occur.

The critical temperature and packing fraction are plotted in Fig. \ref{fig5}. As the pressure increases the packing fraction increases while the critical temperature and concentration tend asymptotically to $kT_c/\epsilon\rightarrow0.037$ and $x_c\rightarrow0.757$.

\subsubsection{$\epsilon^*_{BB}=0,\quad \epsilon^*_{AA}=\epsilon^*_{AB}=1$}\label{subsub3}

Let us continue by considering the extreme case where the $B$ patches do not interact: $\epsilon^*_{BB}=0$. The mixture is completely miscible at pressures above $p_c^{(3)}$. Slightly below $p_c^{(3)}$ the $Tx$ phase diagrams (an example is plotted in panel ($d_1$) of Fig. \ref{fig3}) are similar to those of the reference mixture ($2_A-3_A$): The $LV$ phase transition shifts from $x=0$ ($3_A$ fluid) to finite values of the composition. The differences appear at lower temperatures. As a result of the absence of bonding between $2_B$ particles, there is no condensation when the fraction of $2_B$ particles in the mixture is sufficiently large. When the temperature decreases, at any value of the pressure, the percolation line tends asymptotically to $x\rightarrow0.75$ (with $X_A\rightarrow0$ and $X_B\rightarrow0.5$) rather than $x\rightarrow1$ (with $X_A\rightarrow0$ and $X_B\rightarrow0$) as in the reference mixture. Differences in the phase diagrams are noticeable as the pressure is further reduced (see panel ($d_2$) in Fig. \ref{fig3}). Near $x=0$, there is a $LV$ transition similar to the transition of the reference mixture, but as the temperature is lowered, the two-phase region changes drastically and ends in a lower critical point. Topologically, this is still a limiting case of type I mixtures ($pT$ projection is depicted in Fig. \ref{fig4}). The critical temperature approaches zero when the pressure vanishes, as in the reference mixture, but the critical packing fraction tends asymptotically to $\eta_c=0.0091$ (see Fig. \ref{fig5}) and $x_c\rightarrow0.699$ rather than $\eta_c\rightarrow0$ and $x_c\rightarrow1$. Thus, there is no 'empty' liquid regime when the $B$ patches do not interact. The difference between the $AA$ and $BB$ bonding energies drives demixing between two network fluids, which preempts the $LV$ phase transition at low pressures. 

\subsubsection{$0<\epsilon^*_{BB}<1,\quad \epsilon^*_{AA}=\epsilon^*_{AB}=1$}\label{subsub2}

\begin{figure}
\epsfig{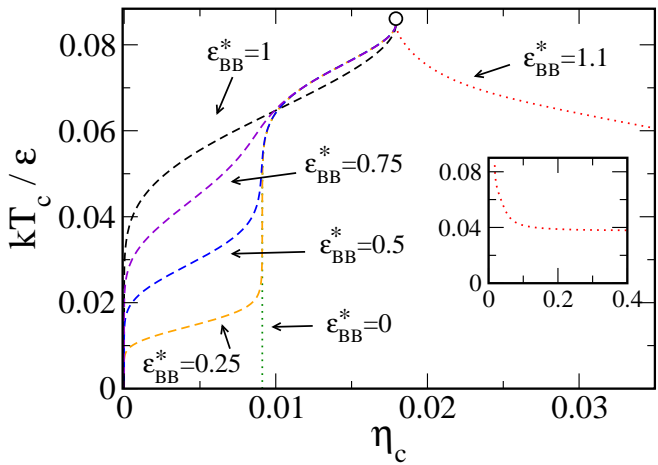}
\caption{Scaled-critical temperature vs critical packing fraction of mixtures with $\epsilon^*_{AA}=\epsilon^*_{AB}=1$. The open circle denotes the critical point of the pure $3_A$ fluid. The inset is a zoom of the case $\epsilon_{BB}^*=1.1$ where $\eta_c\rightarrow1$.}
\label{fig5}
\end{figure}

These mixtures retain features of the previous two classes, those with $\epsilon^*_{BB}=1$ and those with $\epsilon^*_{BB}=0$. An example of the temperature-composition phase diagrams at constant pressure is shown in the third row of Fig. \ref{fig3} for $\epsilon^*_{BB}=0.5$, examples of the $pT$ projections are depicted in Fig. \ref{fig4} and the critical properties are illustrated in Fig. \ref{fig5}. 

At intermediate temperatures, the $BB$ interactions are frozen (in the sense that the probability of $BB$ bonding is much smaller than the probability of $AB$ or $AA$ bonding) and the mixture behaves as the mixture with $\epsilon_{BB}=0$. Eventually, at low enough temperatures, all bonds are established and the mixture behaves as the reference $2_A-3_A$ mixture. In particular, the 'empty' liquid regime ($T_c\rightarrow0$ and $\eta_c\rightarrow0$ as the pressure vanishes) is observed. In all cases these mixtures are limiting cases of type I phase equilibria where only one component undergoes $LV$ condensation.

\subsection{$2_B-3_A$ mixtures: $\epsilon_{AB}=\epsilon_{BB}$}

We continue the analysis by setting $\epsilon^*_{AB}=\epsilon^*_{BB}=1$ and varying the bonding energy of $AA$ bonds.

\subsubsection{$\epsilon^*_{AA}>\epsilon,\quad \epsilon^*_{AB}=\epsilon^*_{BB}=1$}\label{subsub4}

\begin{figure*}
\epsfig{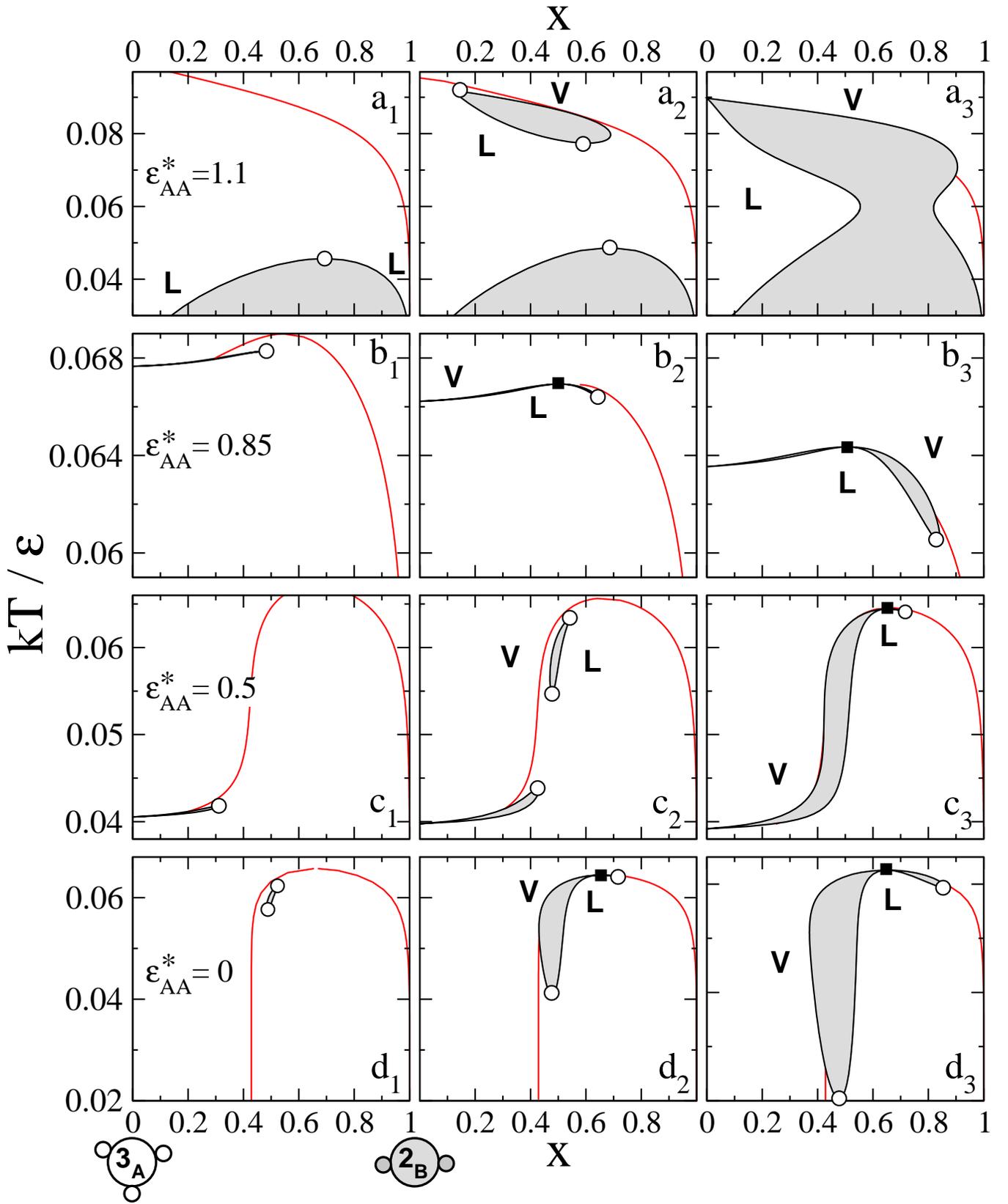}
\caption{Phase diagrams in the scaled-temperature vs composition ($x=x^{(1)}$) plane at constant pressure of $2_B-3_A$ binary mixtures with $\epsilon^*_{AB}=\epsilon^*_{BB}=1$. First row: $\epsilon^*_{AA}=1.1$, pressures: $p^*=pv_s/\epsilon=5.24\times10^{-4}$ ($a_1$), $p^*=3.14\times10^{-4}$ ($a_2$), $p^*=p^*_1=1.05\times10^{-4}$ ($a_3$). Second row: $\epsilon^*_{AA}=0.85$, pressures: $p^*=5.24\times10^{-5}$ ($b_1$), $p^*=3.14\times10^{-5}$ ($b_2$),  $p^*=1.31\times10^{-5}$ ($b_3$). Third row: $\epsilon^*_{AA}0.5$, pressures: $p^*=4.19\times10^{-5}$ ($c_1$), $p^*=2.83\times10^{-5}$ ($c_2$), $p^*=2.09\times10^{-5}$ ($c_3$). Fourth row: $\epsilon^*_{AA}=0$, pressures: $p^*=2.88\times10^{-5}$ ($d_1$), $p^*=2.09\times10^{-5}$ ($d_2$), $p^*=p_2^*=1.05\times10^{-5}$ ($d_3$).} 
\label{fig6}
\end{figure*}

We selected $\epsilon^*_{AA}=1.1$ to illustrate the behaviour of this class of mixtures. Temperature-composition phase diagrams at different pressures are plotted in Fig. \ref{fig6} ($a_1$) $pv_s/\epsilon=5.24\times10^{-4}$ (well above the critical pressure of the pure substance $p_c^{(3)}$), ($a_2$) $pv_s/\epsilon=3.142\times10^{-4}$ (sightly above $p_c^{(3)}$), and ($a_3$) $p_1^*=pv_s/\epsilon=1.05\times10^{-4}$ (sightly below $p_c^{(3)}$). The behaviour is similar to the mixture with $\epsilon^*_{BB}=1.1$ and $\epsilon^*_{AB}=\epsilon^*_{AA}=1$, analysed in Sec. \ref{subsub1}. In addition to the $LV$ condensation, there is a $LL$ demixing region at low temperatures for any value of pressure. Well above $p_c^{(3)}$ the $LL$ demixing region is bounded by an upper critical point, Fig. \ref{fig6} ($a_1$). As $p_c^{(3)}$ is approached from above, the $LV$ condensation appears in the form of closed loops of immiscibility ($a_2$). By further reducing the pressure, the two-phase regions grow and eventually merge giving rise to a very large demixing region, which intersects the $x=0$ axis at the temperature of the $LV$ transition of the pure component ($a_3$). The percolation analysis shows that liquid phases are always network fluids while the vapour phase may or may not be percolated. Closed loops of immiscibility, when present, are entirely percolated.

The origin of the $LL$ demixing at low temperatures is the same as that observed in the mixtures with $\epsilon^*_{BB}=1.1$ and $\epsilon^*_{AB}=\epsilon^*_{AA}=1$. As the temperature is reduced the fractions of unbonded sites, $X_A$ and $X_B$, decrease and there is demixing between a phase where the $AA$ bonds are predominant (small values of composition) and another where the $BB$ bonds predominate (large values of composition). $AB$ bonds are penalised as they reduce the probability of $AA$ bonding, which is energetically favourable. 

This mixture is topologically classified as a limiting case of type III phase equilibria (see the $pT$ projection in Fig. \ref{fig7} (a)) where only one fluid undergoes $LV$ condensation. The critical line starts at the critical point of the pure $3_A$ fluid, exhibits a pressure maximum followed by a pressure minimum at lower temperatures and it continues with negative slope to high pressures, where it is expected to intersect the solid phase lines. The critical density increases monotonically as the pressure is reduced (a representation of $T_c$ vs $\eta_c$ is shown in Fig. \ref{fig8} (a)). Topological changes are not expected for other values of $\epsilon^*_{AA}>1$, although minor modifications of the critical line may occur. Thus, there is no 'empty' fluid regime in this class of mixtures, as it is preempted by $LL$ demixing at low pressures.

\subsubsection{$0<\epsilon^*_{AA}<1,\quad \epsilon^*_{AB}=\epsilon^*_{BB}=1$}\label{subsub5}

In this class of mixtures it is entropically and energetically favourable to form of $AB$ bonds and therefore these mixtures are expected to exhibit 'empty' fluid behaviour. Let us start by reducing the interaction between $A$ patches to $\epsilon^*_{AA}=0.85$. Temperature-composition phase diagrams at constant pressure are depicted in the second row of Fig. \ref{fig6}. The mixture is completely miscible at pressures above $p_c^{(3)}$. By reducing the pressure below $p_c^{(3)}$ we find that the $LV$ condensation of the pure $3_A$ fluid at the $x=0$ axis is shifted to finite values of the composition and ends in an upper critical point. Panel ($b_1$) of Fig. \ref{fig6} is an example. By contrast to the mixtures investigated previously, the slope of the binodal in the $Tx$ plane is positive, indicating an increase in the stability of the liquid phase when $2_B$ particles are added to the mixture. The addition of $2_B$ particles reduces the probability of branching (or equivalently the entropy of bonding) and a negative slope of the binodal, signals a decrease in the stability of the liquid phase. Here, however, the loss in the entropy of bonding is overcome by the gain in bonding energy as the $AB$ bonds are energetically favourable. Note that mixtures where $0<\epsilon^*_{BB}<1,\quad \epsilon^*_{AA}=\epsilon^*_{AB}=1$ (analysed in Sec. \ref{subsub2}) are similar but not exactly the same. In that case the bonding energy of the mixture is also minimised by the formation of $AB$ bonds, with respect to $BB$ bonds, but it is higher than the bonding energy of a pure $3_A$ fluid, whereas in the mixture analysed in this section ($\epsilon^*_{AA}<1,\quad \epsilon^*_{AB}=\epsilon^*_{BB}=1$) it is possible to achieve mixed states with bonding energy lower than that of the pure $3_A$ fluid. 

As pressure is further reduced, a negative azeotrope appears (see panel ($b_2$) of Fig. \ref{fig6}). The second $LV$ branch leaving the azeotrope at high composition grows rapidly as the pressure is reduced (panel ($b_3$) in Fig. \ref{fig6}). This $LV$ branch is bounded by a critical point with critical density and critical temperature that decrease continuously towards zero as the pressure vanishes (see Fig. \ref{fig8} (b)). Therefore, as expected, the mixture exhibits 'empty' fluid behaviour.

\begin{figure}
\epsfig{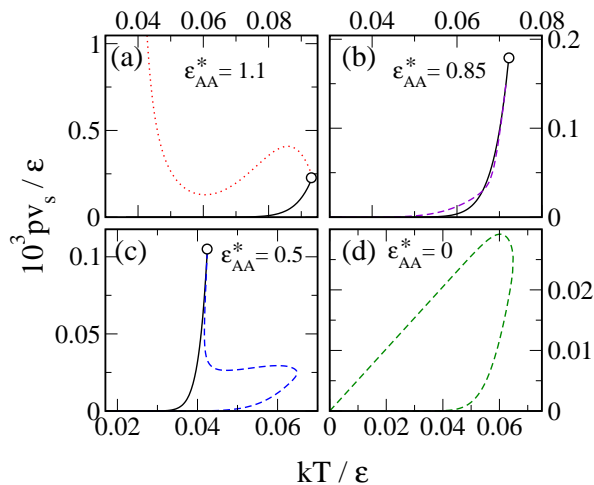}
\caption{Pressure-temperature projections of the phase diagrams of different binary mixtures with $\epsilon^*_{AB}=\epsilon^*_{BB}=1$.}
\label{fig7}
\end{figure}

Topologically this is again a limiting case of type I mixtures with the addition of a negative azeotrope, usually referred to as type I-A phase equilibria \cite{Scott:49} (see the $pT$ projection in panel (b) of Fig. \ref{fig7}).

We proceed by reducing the $AA$ bonding energy to $\epsilon^*_{AA}=0.5$. Three $Tx$ phase diagrams at representative pressures are depicted in the third row of Fig. \ref{fig6}. The mixture is completely miscible at $p>p_c^{(3)}$. Below $p_c^{(3)}$ there is a range of pressures in which the phase diagram consists only of the $LV$ phase transition, shifted from the $x=0$ axis (panel ($c_1$) in Fig. \ref{fig6}). As before, the slope of the binodal is positive. There is a remarkable maximum in the percolation line at temperatures higher than the $LV$ condensation of the pure component. Thus, by increasing the number of particles with two patches it is possible to find reentrant percolation behaviour from a non-percolated to a percolated fluid and back to to a non-percolated fluid. The addition of $2_B$ particles to the mixture reduces branching, but as the $AB$ bonds are energetically favourable, with respect to $AA$ bonds, the probability of branching can, under some circumstances, increase rather than decrease as the composition of $2_B$ particles increases and re-entrant percolation behaviour obtains.

As the pressure is reduced (panel ($c_2$) of Fig. \ref{fig6}) a new two phase region emerges in the form of a closed loop bounded above (below) by an upper (lower) critical point. It is well inside the percolated region and therefore it is a phase transition between two network fluids or percolated states. The new two-phase region is not connected to the $x=0$ axis, clearly showing that the condensation is unrelated to the $LV$ condensation of the pure $3_A$ fluid. As the pressure is further reduced, the distinct two-phase regions merge and, as previously, there is a negative azeotrope (panel ($c_3$) in Fig. \ref{fig6}). Once the azeotrope appears, a new $LV$ branch grows bounded by a critical point, which moves towards lower temperature and density as the pressure vanishes (see Fig. \ref{fig8} ($c$)). The mixture exhibits 'empty' fluid behaviour.

Let us focus now on the $pT$ projection of this mixture, represented in Fig. \ref{fig7} (c). It is still a limiting case of type I-A mixtures (type I with an azeotrope). Interestingly, the critical line leaves the critical point of the pure $3_A$ fluid and goes initially to higher temperatures, a very unusual situation.

\subsubsection{$\epsilon^*_{AA}=0,\quad \epsilon^*_{AB}=\epsilon^*_{BB}=1$}\label{subsub6}

\begin{figure}
\epsfig{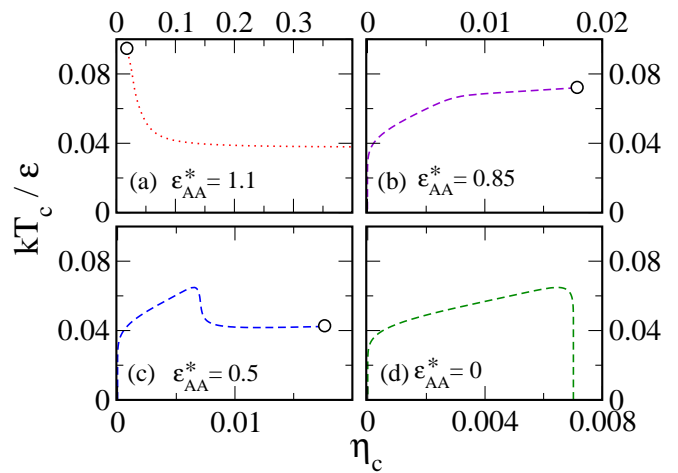}
\caption{Scaled-critical temperature versus critical packing fraction in mixtures with $\epsilon^*_{AB}=\epsilon^*_{BB}=1$.}
\label{fig8}
\end{figure}

The last set of $Tx$ phase diagrams in Fig. \ref{fig6} (fourth row) illustrates the extreme case in which $A$ patches do not interact ($\epsilon^*_{AA}=0$). As a result, none of the two pure fluids undergoes $LV$ condensation. However, the bonding interaction between dissimilar patches ($AB$) is present and drives condensation at pressures below $p^{(+)}\approx2.92\times10^{-5}\epsilon/v_s$. Sightly below $p^{(+)}$ (panel ($d_1$) of Fig. \ref{fig6}) a closed miscibility gap grows inside the percolated region. Well below $p^{(+)}$ (panels ($d_2$) and ($d_3$) of Fig. \ref{fig6}) there is a negative azeotrope with two $LV$ branches bounded by critical points. As the pressure vanishes, both critical temperatures tend to zero, but the critical density behaves differently in each branch (see Fig. \ref{fig8} (d)). The critical point rich in $2_B$ particles exhibits 'empty' fluid behaviour, while the critical point rich in $3_A$ particles tends asymptotically to $\eta_c\rightarrow0.007$.

The behaviour is analogous to the mixture with $\epsilon^*_{AA}=0.5$ with the absence of the $LV$ branch connected to the pure $3_A$ fluid condensation. It is not possible to classify this mixture following van Konynenburg and Scott \cite{Scott:49} as neither of the pure fluids has $LV$ condensation. The critical line in the $pT$ projection (see Fig. \ref{fig7} (d)) forms a closed loop starting and ending at $p\rightarrow0$ and $T\rightarrow0$ where we have found two different solutions. 

\subsection{$2_B-3_A$ mixtures: $\epsilon_{AA}=\epsilon_{BB}$}

\begin{figure}
\epsfig{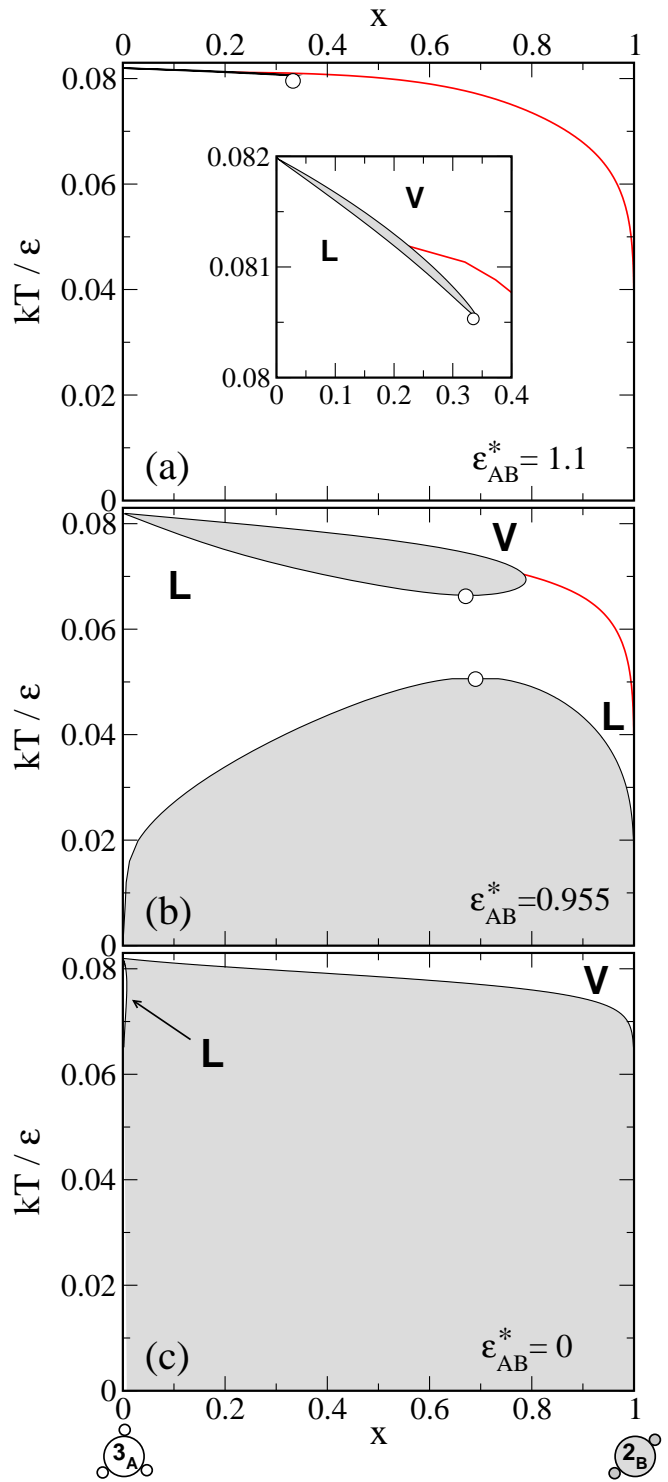}
\caption{Phase diagrams of $2_B-3_A$ binary mixtures in the scaled-temperature vs composition plane at constant pressure $p_1^*=pv_s/\epsilon=1.05\times10^{-4}$. In all cases $\epsilon^*_{AA}=\epsilon^*_{BB}=1$. $\epsilon^*_{AB}=1.1$ ($a$), $\epsilon^*_{AB}=0.955$ ($b$), and $\epsilon^*_{AB}=0$ ($c$). The inset in ($a$) is a zoom of the two-phase region.}
\label{fig9}
\end{figure}

Finally, we analyse mixtures with $\epsilon^*_{AA}=\epsilon^*_{BB}$. Representative examples of $Tx$ phase diagrams at $p_1^*= pv_s/\epsilon=1.05\times10^{-4}$ are plotted in Fig. \ref{fig9}, the pressure-temperature projections are depicted in Fig. \ref{fig10} and the properties of the critical points are shown in Fig. \ref{fig11}.

\subsubsection{$\epsilon^*_{AB}>1,\quad \epsilon^*_{AA}=\epsilon^*_{BB}$ \label{subsub7}}

In this case $AB$ bonds are energetically favourable with respect to both $AA$ or $BB$ bonds, and therefore there is no $LL$ demixing at low temperatures. 

\begin{figure}
\epsfig{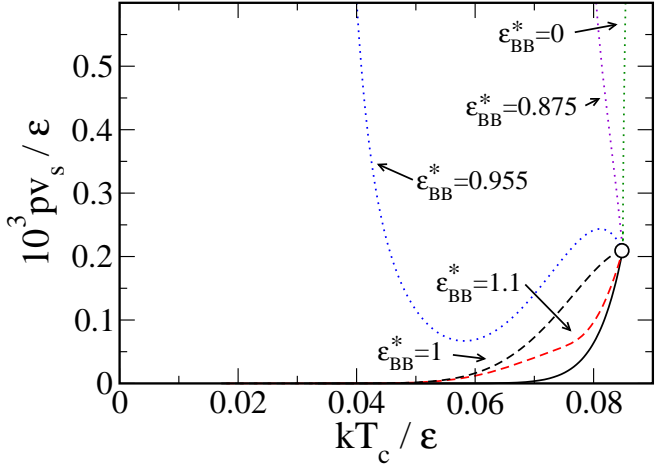}
\caption{Pressure-temperature projections of the phase diagrams of binary mixtures with $\epsilon^*_{AA}=\epsilon^*_{BB}=1$.}
\label{fig10}
\end{figure}

\begin{figure}
\epsfig{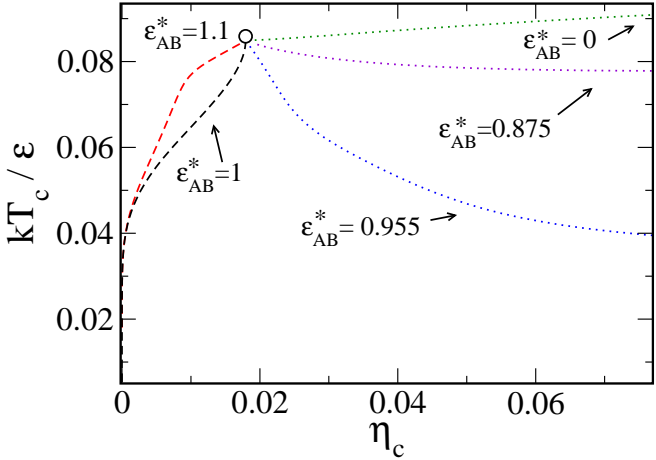}
\caption{Scaled-critical temperature vs critical packing fraction of binary mixtures with $\epsilon^*_{AA}=\epsilon^*_{BB}=1$. In all cases with $\epsilon_{AB}^*<1$ the critical packing fraction tends to $\eta_c\rightarrow1$.}
\label{fig11}
\end{figure}

The $Tx$ phase diagrams are qualitatively the same as those of the reference $2_A-3_A$ mixture. An example is plotted in panel ($a$) of Fig. \ref{fig9} for $\epsilon_{AB}^*=1.1$ at $p_1^*$. It should be compared to the phase diagram of the $2_A-3_A$ reference mixture depicted in panel ($b_1$) of Fig. \ref{fig3}. At constant pressure, the $LV$ two-phase region shrinks as the strength of the interaction between dissimilar patches increases, and the critical point moves towards higher temperature and lower composition. These mixtures are limiting cases of type I (Fig. \ref{fig10}), and, as expected, the critical packing fraction and the critical temperature tend to zero as the pressure vanishes (Fig. \ref{fig11}).  In other words, the mixture exhibits 'empty' fluid behaviour.

Changes in the topology of the phase diagram (as the appearance of negative azeotropes) may arise if the bonding interaction between dissimilar patches is much stronger than the interaction between identical patches: $\epsilon^*_{AB}/\epsilon^*_{\alpha\alpha}>>1,\;\alpha=A,B$. Nevertheless, as far as the critical properties at low pressure are concerned, no changes are expected to occur.

\subsubsection{$0\le\epsilon^*_{AB}<1,\quad \epsilon^*_{AA}=\epsilon^*_{BB}$ \label{subsub8}}

A small decrease in the $AB$ bonding interaction drives phase separation at low temperature, preempting the 'empty' fluid regime. We have plotted two phase diagrams at the same pressure ($p_1^*$) in Fig. \ref{fig9} for: $\epsilon^*_{AB}=0.955$ in panel ($b$) and $\epsilon^*_{AB}=0$ in panel ($c$).  For bonding strengths $\epsilon_{AB}^*\lesssim1$ there are two regions of phase separation ($LL$ demixing bounded by an upper critical point at low temperatures, and $LV$ condensation bounded by a lower critical point at high temperatures) at pressures below the critical pressure of the pure $3_A$ fluid ($b$). When $\epsilon_{AB}^*<<1$, both regions merge into a single two-phase region without critical points below $p_c^{(3)}$, see for example panel ($c$). In both cases, at pressures above $p_c^{(3)}$, $LL$ demixing is always present bounded by an upper critical point. These mixtures are limiting cases of type III mixtures (see Fig. \ref{fig10}). The strength of the $AB$ interaction leads to slight changes of the critical line. Note, for example, the slope of the critical line at high pressures: Positive when $\epsilon^*_{AB}<<1$ and negative when $\epsilon_{AB}^*\lesssim1$.

\section{Conclusions \label{conclusions}}

We have carried out a systematic analysis of binary mixtures of patchy particles with $2$ patches of type $B$ and $3$ patches of type $A$, focusing on the 'empty' fluid behaviour discovered in one of these mixtures \cite{PhysRevLett.97.168301,bianchi2008}. We have found a rich phase behaviour with closed miscibility gaps, liquid-liquid demixing, negative azeotropes and, from a topological point of view, three different types of mixtures: Type I, I-A and III. The main results are summarised in Table \ref{table1}. 

\begin{table}[h]
\begin{center}
\begin{tabular}{|c|c|c|c|}
\hline
mixture & empty fluids & type & section \\
\hline
$\epsilon^*_{BB}>1$ & no & III & \ref{subsub1}\\
\hline
$0<\epsilon^*_{BB}<1$ & yes & I & \ref{subsub2}\\
\hline
$\epsilon^*_{BB}=0$ & no & I & \ref{subsub3}\\
\hline
$\epsilon^*_{AA}>1$ & no & III & \ref{subsub4}\\
\hline
$0<\epsilon^*_{AA}<1$ & yes & I-A & \ref{subsub5}\\
\hline
$\epsilon^*_{AA}=0$ & yes & -- & \ref{subsub6}\\
\hline
$\epsilon^*_{AB}>1$ & yes & I & \ref{subsub7}\\
\hline
$0<\epsilon^*_{AB}<1$ & no & III & \ref{subsub8}\\
\hline
$\epsilon^*_{AB}=0$ & no & III & \ref{subsub8}\\
\hline
\end{tabular}
\caption{Summary of the phase behaviour of $2_B-3_A$ mixtures: The bonding energy that characterizes the mixture (the other two are set to one), the stability of 'empty' fluids and the topological classification of the mixture phase diagram.}\label{table1}
\end{center}
\vspace{-0.6cm}
\end{table}

We have found that liquid-liquid demixing, at low temperatures and pressures, is the primary mechanism that preempts the 'empty' fluid regime in some of these mixtures. $LL$ demixing is always present when unlike ($AB$) bonds are weaker than at least one of the other two bonds between particles of the same species ($AA$ or $BB$ bonds). These mixtures are limiting cases of type III. Within this class the detailed phase behaviour depends on the values of the bonding energies. For example, the slope of the critical line in the $pT$ projection, the appearance of closed miscibility gaps or the asymptotic values of the critical temperature and critical concentration as the pressure increases, are controlled by the specific values of the bonding energies.

The mixtures exhibit 'empty' fluid behaviour if the $AB$ bonds are stronger than at least one of the other two. These mixtures are limiting cases of type I or type I-A (type I with the addition of a negative azeotrope)\footnote{There is one exception, when $\epsilon_{AA}=0$ none of the pure fluids undergoes $LV$ condensation, and therefore it is not possible to classify this mixture within the standard classification of van Konynenburg and Scott \cite{Scott:49}.}. The most interesting case occurs in mixtures with bonding strengths $\epsilon^*_{BB}=\epsilon^*_{AB}$ and $\epsilon^*_{AA}<<1$. In these mixtures there is a negative azeotrope and the critical line goes initially to temperatures above that of the pure $3_A$ fluid. 

There is only one case where $AB$ bonds are energetically favourable and the mixture does not exhibit 'empty' fluid behaviour: The mixture with  $\epsilon_{AB}=\epsilon_{AA}$ and $\epsilon_{BB}=0$. In this case the absence of bonding between particles with $2$ patches precludes the 'empty' liquid regime. In fact, near the critical point of the mixture there is coexistence between two network fluids with finite packing fractions as the pressure vanishes. This behaviour is similar to that found previously \cite{mixtures1} for $2_A-5_A$ binary mixtures ({\it i.e} mixtures of particles with $2$ and $5$ identical patches) where the entropy of bonding drives liquid-liquid demixing preempting the liquid-vapor phase transition.

In a previous study \cite{mixtures1} we found that the driving force for the topological change in the phase diagram of mixtures with a single bonding energy 
is the difference in bonding entropy associated with bonds between like- and unlike- particles. It is desirable to address this question in more detail by investigating mixtures of particles with different diameters, and/or, different types of bonding sites in such a manner that both fluids are constrained to have similar values of the critical pressure and temperature. In that context, the present study is a step towards elucidating the competition between bonding entropy and bonding energy in the determination of the topology of the phase diagrams of binary mixtures.  

Binary mixtures of particles with different types of bonding sites are also crucial to the investigation of a wide range of structured fluids, with different macroscopic properties, including stable bigel phases \cite{goyal:064511,*B907873H}. This is an immediate goal that will be addressed in future work. 

Finally, a word about the absolute stability of the fluid phases reported in this paper. Although at low pressures and low (but finite) temperatures we expect the fluid to be absolutely stable it is clear that solid phases will intervene and preempt some of the high pressure and/or low temperature features of the phase diagrams reported here. The calculation of solid phase diagrams is much more complicated but progress has been reported recently, including ground state analysis of binary mixtures of low-functionality patchy particles in two dimensions \cite{0953-8984-22-10-104105,B614955C}.

\section{Acknowledgments}

This work has been supported, in part, by the Portuguese Foundation for Science and Technology (FCT) through Contracts Nos. POCTI/ISFL/2/618 and PTDC/FIS/098254/2008, by the R$\&$D Programme of Activities (Comunidad de Madrid, Spain) MODELICO-CM/S2009ESP-1691, and by the Spanish Ministry of Education through grant FIS2008-05865-C02-02. D. de las Heras is supported by the Spanish Ministry of Education through contract No. EX2009-0121.

\end{document}